\documentclass[aps,pre,twocolumn,groupedaddress,showpacs,superscriptaddress,amssymb,amsmath,longbibliography]{revtex4-1}
\usepackage{physics}
\usepackage{verbatim}
\usepackage{graphicx}
\usepackage{amsmath}
\usepackage{amssymb}
\usepackage{amsfonts}
\usepackage{color}
\usepackage{bm}      
\usepackage{verbatim}
\usepackage{placeins}
\usepackage[mathscr]{eucal}
\usepackage[colorlinks, linkcolor=red,citecolor=blue,urlcolor=blue]{hyperref}
\usepackage[normalem]{ulem}
\usepackage[all]{hypcap}
\usepackage{multirow}
\usepackage[dvipsnames,usenames,table]{xcolor}
\usepackage{dcolumn}
\usepackage{hyperref}
\usepackage{bm}
\usepackage{epsf}
\usepackage{subfigure}
\usepackage{amsfonts}
\usepackage{epstopdf}
\newcommand{\be}{\begin{equation}}
\newcommand{\ee}{\end{equation}}
\newcommand{\bea}{\begin{eqnarray}}
\newcommand{\eea}{\end{eqnarray}}
\definecolor{mblue}{rgb}{0.0,0.45,0.74}
\usepackage{hyperref}
\usepackage{amsfonts}
\usepackage{tikz}
\usepackage{pgfplots}

\begin{document}








\title{Hyper-acceleration of quantum thermalization dynamics by bypassing long-lived coherences: An analytical treatment}


\author{Felix Ivander}
\affiliation{Chemical Physics Theory Group, Department of Chemistry and Centre for Quantum Information and Quantum Control,
University of Toronto, 80 Saint George St., Toronto, Ontario, M5S 3H6, Canada}
\author{Nicholas Anto-Sztrikacs}
\affiliation{Department of Physics, 60 Saint George St., University of Toronto, Toronto, Ontario, Canada M5S 1A7}

\author{Dvira Segal}
\email{dvira.segal@utoronto.ca}
\affiliation{Chemical Physics Theory Group, Department of Chemistry and Centre for Quantum Information and Quantum Control,
University of Toronto, 80 Saint George St., Toronto, Ontario, M5S 3H6, Canada}
\affiliation{Department of Physics, 60 Saint George St., University of Toronto, Toronto, Ontario, Canada M5S 1A7}

\date{\today}

\begin{abstract}
We develop a perturbative technique for solving Markovian quantum dissipative dynamics,
with the perturbation parameter being a small gap in the eigenspectrum.
As an example, we apply the technique and straightforwardly obtain \textit{analytically} the dynamics of a three-level system with 
quasidegenerate excited states, where quantum coherences persist for very long times, proportional to the inverse of the energy splitting squared.
We then show how to \textit{bypass} this long-lived coherent dynamics 
and accelerate the relaxation to thermal equilibration in a \textit{hyper-exponential} manner, a
Markovian \textit{quantum-assisted} Mpemba-like effect.
This hyper-acceleration of the equilibration process manifests if the initial state is 
carefully prepared, such that its coherences precisely store 
the amount of population relaxing from the initial condition to the equilibrium state.
Our analytical method for solving quantum dissipative dynamics readily provides equilibration timescales, and as such it reveals how coherent and incoherent effects 
interlace in the dynamics. It further advices on how to accelerate relaxation processes, which is desirable  when long-lived quantum coherences stagnate dynamics.
\end{abstract}

\maketitle


\section{Introduction}

The survival, manifestation, and control of quantum effects in noisy environments is a foundational problem in quantum dynamics \cite{BPBook} with applications ranging from biology, e.g., in the processes of vision \cite{V1,V2} and photosynthesis \cite{Olaya,Kassal16, Kassal21,QB,Greg,Dwayne17},
to quantum technologies including quantum computing \cite{chuang},  sensing \cite{Paola}, metrology \cite{VmodelNJP}, and quantum thermodynamics \cite{QAR,db1,db2,SW1,SW2,SW3,DBc2,QTMC1,QTMC2,COHQAR1,COHQAR2,COHQAR3,fuel,friction}. These references are only examples of an extensive literature.
%
From a theoretical standpoint, manifolds with quasidegenerate levels, such as the V and the $\Lambda$ models,  
are archetypal in the ongoing efforts to understand the impact of quantum coherences on dissipative dynamics. This owes to their simple yet rich physics, and to the fact that such level-schemes show up as limiting configurations in natural systems, e.g., in atoms with hyperfine levels e.g., Rubidium \cite{EIT,expe1,VExp1}, Cesium \cite{expe2}, and Sodium vapor \cite{expe3}. Furthermore, such level-schemes also serve as building blocks to more elaborate systems including those supporting photosynthetic energy transfer \cite{ScullyPNAS2013} and multilevel quantum thermal machines \cite{MK,VmodelNJP}. 

The behavior of quantum coherences in quasidegenerate level-schemes have been extensively studied in the framework of quantum optics \cite{PhysRevA.46.373,PhysRevA.47.2186,Agarwal1999,CPT,Li_2000,Scully,EIT,EITexp,LWI5,AO,KIFFNER201085,Vmodelcorr,VmodelNJP} and quantum biology \cite{b1,b2,b3,b5,Dodin_2016,Timur14,JCao}, with varying conclusions on whether coherences  are advantageous to performance, or otherwise. Notably, Tscherbul and Brumer \cite{Timur14} showed that in the V model the lifetime of coherent dynamics scales inversely to the level degeneracy-squared, an effect elaborated on in Refs. \cite{Tscherbul_2014,Tscherbul_2015,Dodin_2016, dodin_quantum_2016,
Dodin_2021,Tscherbul_2022, Merkli2015}. 
In recent studies, it was further demonstrated that quantum coherences can be deleterious to steady state heat transport  \cite{MK,NJP}. In such applications, long-lived coherent dynamics are not desirable. Accordingly, a method to bypass slow dynamics en route to equilibrium is desired for the optimal design of quantum devices.

The Mpemba effect is a phenomenon by which, under otherwise identical external conditions, a system starting {\it farther} from equilibrium reaches it  {\it sooner} than a system that starts {\it closer} to equilibrium. 
Anomalous cooling of water is a classical example, reported as early as the third century B.C. by Aristotle \cite{Linden}, but analogous phenomena had also been observed in spin glasses \cite{MpembaSG}, granular gases \cite{MpembaGG1,MpembaGG2,MpembaGG3,MpembaGG4}, nanotube resonators \cite{MpembaNR}, clathrate hydrates \cite{MpembaCH}, colloids \cite{MpembaC}, ultracold atoms \cite{MpembaUC}, polylactide (PLA) \cite{MpembaPLA}, and magnetic alloys \cite{MpembaMA}. 
Nonetheless, even for anomalous relaxation of water, the fundamental mechanism underlying the Mpemba effect is still not fully resolved -- proposals range from supercooling \cite{Supercooling}, minute impurities \cite{Katz,Bednarz}, convection \cite{conv}, evaporation \cite{evp,Kell}, dissolved gases \cite{gas}, to peculiar properties of hydrogen bonds \cite{CGD,WAG,JCTC,PCCP}. 

Recently, the Mpemba effect has been theoretically investigated using the framework of classical and quantum dynamics \cite{Mpemba1,MpembaPNAS,MpembaPRL,MpembaPRX,MpembaNJP,MpembaPRR,Raz}, demonstrating that neither \textit{memory} nor \textit{non-Markovianity} are necessary conditions for the effect. 
A generic mechanism underlying anomalous relaxation (cooling or heating) was proposed in Ref. \cite{MpembaPNAS}, and later verified in a set of carefully-controlled experiments \cite{MpembaC,John22,johnR}: Expressing the dynamics with a linear combination of eigenmodes, the Mpemba effect of anomalous cooling occurs if the initial, high-temperature state of the system has a {\it smaller amplitude} of its slowest-decaying mode, than a state prepared at a lower temperature. An analogous explanation leads to the ``inverse Mpemba effect" of anomalous heating. 

The focus of our study are open quantum systems, for which the Mpemba effect offers a mean to modify and control equilibration dynamics by cutting short long-lived coherent dynamics \cite{MpembaPRL}.
The objectives of this paper are twofold. 
First, we devise and exemplify a perturbative method for solving Markovian quantum dissipative dynamics when a small energy splitting in the spectrum can be identified as the perturbative parameter. 
Particularly, the method perfectly fits to solve the recently-developed Unified quantum master equation (QME) \cite{Anton,Gerry}, where the generator of the dynamics is partitioned into two terms, a zeroth order term that averages over small energy splittings, and a term accounting for those close-to-degeneracy contributions. 
Our second objective is to show how to bypass slow dynamics in systems with long-lived coherences via an Mpemba-like quantum effect.
We demonstrate this process analytically with our perturbative method: 
By preparing initial conditions that are {\it orthogonal} to the slowest decaying mode, we hyper-accelerate the relaxation dynamics towards equilibrium and avoid long-lived, stagnating transients. 
We find that these Mpemba initial conditions should have coherences that (i) lie within a certain range and (ii) precisely compensate the difference between the final (equilibrium)  and initial-level populations.  
The mechanism of the Mpemba effect in quantum systems is analogous to the classical one \cite{MpembaPNAS,johnR}. However, in quasi-degenerate levels the Mpemba effect is considerably more dramatic than under classical dynamics due to interlacing unitary and dissipative effects.

This work is organized as follows. In Sec. \ref{S:1Method} we present the Liouvillian eigenvalue estimation technique underpinning the derivation of analytical results.  
We describe the V model and its equations of motion, and solve the dynamics analytically in Sec. \ref{S1:Model}
with some technical details delegated to Appendix A.
We apply the method to investigate the quantum Mpemba effect in Sec. \ref{S2:Mpemba}.
We provide a perspective of our results in the Discussion Section \ref{S:Discussion}, along with presenting  extensions to the V model (Appendix B) and 
the solution to the corresponding classical Mpemba effect (Appendix C).
We conclude in Sec. \ref{S3:Summary}.
\section{Liouvillian eigenvalues perturbative estimation (LEPE) technique}
\label{S:1}

\subsection{Presentation of Method}
\label{S:1Method}

In this section, we introduce our method for approximating the eigenvalues of the Liouvillian superoperator responsible for the Markovian dynamics of an open quantum system. We refer to the method as the Liouvillian eigenvalues perturbative estimation (LEPE) technique. We consider a system's Hamiltonian with an energy spectrum containing some nondegenerate levels, as well as nearly-degenerate levels with gaps characterized by the energy parameter $\Delta$. Here, $\Delta$ is small compared to any other level splitting, as well as to temperature and the inverse of relaxation time of the system \cite{comment3}, thus serving as the perturbative parameter of the problem.

Several recent studies had developed perturbative treatments for solving the dynamics and steady state of open quantum systems, yet to the best of our knowledge, none had targeted the questions and the type of perturbation examined here:
Ref. \cite{Popkov} focused on the spectral properties of the Liouvillian, as we do, yet with the perturbative parameter being the inverse dissipation rate constant. Refs. \cite{Koch14,Koch16} describe in rigor a general perturbative approach, applied to build, in a perturbative fashion, the  {\it steady state} density matrix. In our study, however, given the type of perturbation that we employ, the steady state is in fact trivial (canonical equilibrium), while the perturbative approach serves us to construct the relaxation timescales, which are non-trivial, displaying long-lived coherent dynamics. 

In our model, the system's Hamiltonian $\hat{H}_S$ of rank $N$
is coupled to a heat bath maintained in a thermal-canonical state. 
We assume that  the equation of motion (EOM) for the reduced density matrix of the system, $\sigma(t)$, follows a time-local quantum master equation of the form
\bea
    \dot{\sigma}(t) = -i[\hat{H}_S,\sigma(t)] + \mathcal{D}[\sigma(t)].
    \label{eq:Redfield}
\eea
Here, $\mathcal{D}$ is the superoperator responsible for dissipative dynamics. Such a QME arises, e.g., under the Born-Markov approximation
when the system is weakly coupled to a fast heat bath. Moreover, this form also appears in situations beyond the weak coupling limit: The polaron-transformed QME can be also made time-local \cite{CaoPT,Erik}, as well as the reaction-coordinate QME \cite{RCNick}. Both methods account for system-bath couplings beyond the weak (Born) approximation; the reaction-coordinate QME further includes non-markovian effects \cite{RCmark}.

It is possible to recast Eq. (\ref{eq:Redfield}) in the form of a matrix equation by vectorizing the reduced density matrix, 
putting all components of $\sigma(t)$ into a vector, $\vec{x}(t) = (\sigma_{11}(t),\sigma_{12}(t), \ldots )$.
In principle, the vector has $N^2$ elements and we can describe the method as such.
However, for convenience of later discussion we use the population normalization condition $\sum_i{\sigma_{ii}(t)}=1$, and thus reduce the number of elements by 1.  Doing so we obtain an inhomogeneous EOM,
\bea
    \dot{\vec{x}}(t) = L\vec{x}(t) +\vec{d}.
    \label{eq:MatrixQME}
\eea
The Liouvillian matrix $L$ contains contributions from both the unitary and dissipative parts of the master equation.
The vector $\vec d$ emerges due to the normalization condition  employed. 
The main difference of Eq. (\ref{eq:MatrixQME}) from the classical counterpart is that in classical systems, $\vec{x}$ would be the population (or probability) vector, see Appendix C for the classical case. In contrast, here $\vec{x}$ accounts also for coherences.
We now make an eigenvalue expansion ansatz for each component of this equation, in the form
\bea
    x_{i}(t) = x_{\infty,i} + \sum_{n=1}^{N^2-1} c_{n,i}e^{\lambda_n t}.
    \label{eq:ansatz}
\eea
 $x_{\infty,i}$ is the long-time limit of $x_i(t)$, obtained from the steady state solution of Eq. (\ref{eq:MatrixQME}),
\bea
\vec{x}_{\infty}=-L^{-1}\vec d.
\eea
Back to Eq. (\ref{eq:ansatz}),
we proceed to (i) approximate  the eigenvalues $\lambda_n$ and (ii) relate the expansion coefficients of the $i$th element, $c_{n,i}$, to the initial conditions.

\subsubsection{Approximating the eigenvalues of $L$}

$\lambda_n$ is the $n$th eigenvalue of $L$, determined from the characteristic equation
\bea
    f(\lambda) \equiv \det|\lambda I - L| = 0,
\label{eq:char}
\eea
with $I$ as the identity matrix.
We note that depending on the symmetry of the system's Hamiltonian and how the system couples to the bath,
 one can further reduce the complexity of the problem by working with a vector $\vec x$ of reduced dimensionality, smaller than $N^2-1$.
For example, for a three-level system there are in general eight elements in $\vec x$ 
(we exclude say $\sigma_{11}(t)$ through the conservation of population condition). However, in the V model presented in Sec. \ref{S1:Model}, based on symmetries specific to the model, we manage to reduce the dimensionality of $\vec x$ to three.

The eigenvalues of $L$ are generally difficult to obtain analytically-exactly. We now invoke our working assumption, that the spectrum of $\hat{H}_S$ possesses discrete levels, as well as a manifold of nearly-degenerate levels. We then define a simpler problem as follows.
The matrix $L$ is partitioned, $L = L^{(0)} + L^{(1)}$, with $L^{(0)}$ 
describing the dynamics of the fully-degenerate system and $L^{(1)}$ comprising small degeneracy-breaking perturbations, typically arising from the unitary part of the dynamics. 
To zeroth order in the small energy parameter $\Delta$, the eigenvalues $\lambda$ of $L$ are given by the eigenvalues $\lambda^{(0)}$ of $L^{(0)}$, which we find from the characteristic equation 
\bea
    g(\lambda^{(0)}) \equiv \det|\lambda^{(0)} I - L^{(0)}| = 0.
\eea
It is typically {\it easier} to diagonalize $L^{(0)}$, compared to the full $L$
since it does not depend on small energy splittings. 
This is particularly true if we adopt the Unified QME \cite{Anton}, a completely positive
and trace preserving (CPTP) map, which is also  consistent in a thermodynamical sense \cite{Gerry}.
In that case, $L^{(0)}$ included only the dissipative (classical-like) component of the Liouvillian, which had been analyzed analytically in different models, e.g., in the context of random walks  \cite{Klafter}.
The contribution of $L^{(1)}$ may be however crucial to the dynamics; 
we now assume that the eigenvalues of $L$, to lowest order in the small energy splitting
$\Delta$ are given by $\lambda_n \approx \lambda_n^{(0)} + \delta_n$ \cite{comment},
where $\delta_n$ are small corrections to the eigenvalues of $L^{(0)}$,  $|\delta_n| \ll |\lambda_n^{(0)}|$. 
These corrections are obtained by solving Eq. (\ref{eq:char}) under the ansatz
\bea
    f(\lambda^{(0)}_n + \delta_n) = 0,
    \label{eq:flam}
\eea
while keeping only leading order $\Delta$ terms in $\delta$.
Note that because we reduced the dimensionality of the Liouvillian by enforcing the normalization condition in Eq. (\ref{eq:MatrixQME}), the real part of all the eigenvalues $\lambda_n$ is negative.  
We exemplify this process in Sec. \ref{S1:Model} on the V model. 
There, instead of tackling a cubic equation to resolve the eigenvalues of $L$, we end up solving a simpler, quadratic equation for $L^{(0)}$. Assuming small corrections, we find the roots of Eq. (\ref{eq:flam}) with algebraic manipulations and build the eigenvalues of $L$.

\subsubsection{Approximating the expansion coefficients}

Our next task is  to express the coefficients $c_{n,i}$ in terms of the initial conditions. Recall that $n$ indicates the mode and $i$ identifies the element of the reduced density matrix.
First, we note that from the ansatz, Eq. (\ref{eq:ansatz}), we get
\bea
    \frac{d^k}{dt^k}x_i(t)\Big|_{t=0} = \sum_{n=1}^{N^2-1} c_{n,i}\lambda_n^k,  \,\,\,\,\, k>1,
\eea
which we reformulate as a matrix-vector operation,
\bea
\begin{bmatrix}
x_i(0)\\
\dot{x}_i(0)\\
\ddot{x}_{i}(0)\\
\vdots
\end{bmatrix}=\underbrace{\begin{bmatrix}
1 & 1 & 1 & ... \\
\lambda_1 & \lambda_2 & \lambda_3 & ...\\
\lambda_1^2 & \lambda_2^2 & \lambda_3^2 & ...\\
\vdots & \vdots & \vdots & \ddots \\
\end{bmatrix}}_{{\Lambda}}\underbrace
{\begin{bmatrix}
c_{1,i}\\
c_{2,i}\\
c_{3,i}\\
\vdots
\end{bmatrix}}_{\vec{c}^{(i)}}
+\underbrace{
{\begin{bmatrix}
x_{\infty,i}\\
0\\
0\\
\vdots
\end{bmatrix}}}_{{\vec{x}^{(i)}_{\infty}}}.
\nonumber\\
\label{eq:Omega}
\eea
The derivatives of $x_i(t)$ at time zero (left hand side of the above expression) can also be expressed in terms of the initial conditions $\vec{x}(0)$.
Specifically, we write down the sequence of relations, 
\bea
\vec x(0)&=&\vec x(0),
\label{eq:S1} 
\\
\dot {\vec x}(0) &=& L \vec x(0) + \vec {d}, 
\label{eq:S2}
\\
\ddot {\vec x}(0) &=& L^2 \vec x(0)+L\vec d,
\label{eq:S3}
\\
\dddot {\vec x}(0) &=& L^3 \vec x(0)+L^2\vec d,
\eea
and so on,
allowing us to express the left hand side of Eq. (\ref{eq:Omega})
as a matrix operation $B^{(i)}$ on  $\vec x(0)$, in addition to a constant contribution. 
Explicitly, the first row of $B^{(i)}$ is made of  zero elements, except a single `1' in the $i$th place [(Eq. (\ref{eq:S1})].
The second row of $B^{(i)}$ is constructed from the $i$th row in the relation $L\vec x(0)$ [Eq. (\ref{eq:S2})]; the third 
row  is constructed from the $i$th row of $L^2\vec x(0)$ [(Eq. (\ref{eq:S3})], and so on. Similarly, we collect constant terms
and build the vector $\vec{v}^{(i)}=(0, \vec d_i, (L\vec d)_i, (L^2\vec d)_i, ...)^{T}$,
where $\vec d_i$ refers to the $i$th element in the vector $\vec d$. These relations are obtained by taking the $n$th derivative of Eq. (\ref{eq:MatrixQME}) at time zero and iteratively substituting occurrences of $\dot{\vec{x}}(t)$ again with Eq. (\ref{eq:MatrixQME}). Appendix A demonstrates this process for the V model.
Altogether we establish a linear relationship between initial conditions on the density matrix and the coefficients of the different modes,
\bea
B^{(i)}\vec{x}(0) + \vec{v}^{(i)} = \Lambda\vec{c}^{(i)} +\vec x^{(i)}_{\infty}.
\label{eq:BB}
\eea
Note that the matrix $B^{(i)}$, as well as the  vectors $\vec x_{\infty}^{(i)}$ and $\vec v^{(i)}$ are in general 
distinct for each element of the reduced density matrix, $x_i(t)$. 
To be consistent with the lowest-order expansion in $\Delta$, one should solve Eq. (\ref{eq:BB}) in the $\Delta\to 0$ limit, which further simplifies the problem. 

We summarize the relationship (\ref{eq:BB}) as follows:

(i) This algebraic equation displays the relation between the initial condition on the reduced density matrix $\vec x(0)$ and the expansion coefficients of the different eigenmodes of the dynamics $\vec c^{(i)}$. In common approach to dynamics, one prepares the reduced density matrix,  typically finding that all expansion coefficients are nonzero. However, as we discuss in Sec. \ref{S1:Model}, an acceleration of the dynamics can be realized by engineering an initial condition in which the coefficient of the slowest mode is null, or sufficiently small,
thus resulting in a Mpemba-like anomalous relaxation dynamics.

(ii) The dimension of the matrix $\Lambda$ (and $B^{(i)}$) is equal to the number of eigenvalues that one needs to solve for, that is the dimension of $L$.

(iii) The LEPE method is general, but it is particularly useful for solving the dynamics of  systems with nearly-degenerate levels. 


In what follows, we exemplify the LEPE approach and study the dynamics of population and coherences in the V model. 
An extension of the method to treat systems with multiple manifolds of quasi-degenerate levels is straightforward. Additionally, consecutive applications of this protocol can be performed to account for multiple small perturbative parameters. 

\subsection{Example: The V model}
\label{S1:Model}

\subsubsection{Hamiltonian and equations of motion}

Long-lived environmentally-induced quantum coherences are best exemplified with the V model: The excited states of the V model are quasi degenerate, and they are both accessible through bath-induced transitions from the ground state. Analytical solutions for the dynamics of the V model were presented by Brumer and coworkers, see e.g., Refs. \cite{Tscherbul_2014,Tscherbul_2015,Dodin_2016}.
It was shown that the lifetime of coherent dynamics scales {\it inversely} with the level degeneracy squared. Here, we show that the LEPE method of Sec. \ref{S:1Method} accurately and without difficulty solves this problem by going around the need to solve a cubic equation. 

The Hamiltonian of the V system is given by 
\bea
\hat{H}_S=(\nu-\Delta)|2\rangle\langle2|+\nu|3\rangle\langle3|.
\label{eq:Hs}
\eea 
We adopt here natural units, $\hbar=1$, $k_B\equiv1$, and further work in the limit of $\Delta\ll\nu$; thus, levels $|2\rangle$ and $|3\rangle$ are nearly degenerate. Transitions between ground and excited states of the system are enacted by a heat bath at temperature $T$, given by the Hamiltonian,
\bea
\hat{H}_{B}=
\sum_{j}\omega_{j}\hat{b}_{j}^\dagger\hat{b}_{j}. 
\label{eq:HB}
\eea %
Here, $\hat{b}_{j}^\dagger$ ($\hat{b}_{j}$) is the creation (annihilation) bosonic operator of a mode $j$ of frequency $\omega_{j}$.
The system-bath interaction Hamiltonian is given in a bipartite form, with a system operator $\hat{S}$ coupled to a bath operator $\hat B$,
\bea
\hat{H}_{SB}=  \hat{S}\otimes\hat{B};   \,\,\,\,\,\
\hat{B}=\sum_jg_{j}(\hat{b}_{j}^\dagger+\hat{b}_{j}). 
\label{eq:Vm}
\eea
 $g_{j}$ describes the system-bath coupling energy between mode $j$ in the bath and the system. The bath excites the following 
 transitions:
\bea
\hat{S}=|1\rangle\langle2|&+&
|1\rangle\langle3|+h.c. 
\label{eq:S} 
\eea
Here, $h.c.$ is a hermitian conjugate. Overall, the total Hamiltonian is given by
%
$\hat{H}=\hat{H}_S+\hat{H}_{B}+\hat{H}_{SB}$,
%
and our objective is to study the reduced dynamics of the V system. 


We adopt the Unified QME approach, a simplified Redfield treatment that is a CPTP map \cite{Anton}, and further satisfies the fluctuation symmetry \cite{Gerry}.
In this approach, quasidegenerate levels are clustered into the same value when evaluating the dissipative dynamics; the coherent part of the evolution accounts for small energy splitting. For the V model, we receive the following EOM \cite{NJP},
\bea
\dot{\sigma}_{32}(t)&=& -i\Delta\sigma_{32}(t)
- k\sigma_{32}(t)
\nonumber\\
&-& \frac{1}{2} 
k \left[ \sigma_{22}(t)+  \sigma_{33}(t) \right] + 
k e^{-\beta\nu} \sigma_{11}(t)
\label{eq:s23}, 
\\
\dot{\sigma}_{22}(t)&=&
- k \sigma_{22}(t) 
+ ke^{-\beta \nu}\sigma_{11}(t) - k \sigma_{32}^R(t),
\label{eq:s22}
\\
\dot{\sigma}_{33}(t)&=&
-k \sigma_{33}(t) + 
ke^{-\beta \nu} \sigma_{11}(t) -  
k \sigma_{32}^R(t),
\label{eq:s33}
\eea
along with the population conservation condition, $\sum_i \sigma_{ii}(t)=1$. Here and to follow, $\sigma_{32}^R(t)\equiv\text{Re}\sigma_{32}(t)$ and $\sigma_{32}^I(t)\equiv\text{Im}\sigma_{32}(t)$. Note that in the V model with
the coupling operator Eq. (\ref{eq:S}), 
the elements $\sigma_{13}$ and $\sigma_{12}$ are exactly decoupled from this dynamics at the level of the Redfield QME; we did not invoke here the secular approximation. Furthermore, 
in the Unified QME approach, which builds on the Redfield formalism, rate constants between either excited levels and the ground state are assumed equal,  $k_{3\to1}\approx k_{2\to1}\equiv k$. This is because once $\Delta \ll \nu$  we cluster levels in each quasi-degenerate levels manifold to a single value \cite{Anton,Gerry}, as the lowest-order contribution of energy differences become negligible.

For the bosonic-bath model with ${\hat H}_{SB}$
as presented above, the rate constants are given by
a product of the spectral density function 
$J(\omega)\equiv\sum_{j}\pi g_j^2\delta(\omega-\omega_j)$ and the  Bose Einstein distribution function, $n_B(\nu) =\left(e^{\beta \nu} -1\right)^{-1}$,
\bea
k(\omega)=2 J(\omega) [n_B(\omega)+1].
\label{eq:kk}
\eea
However, motivated by the physical setup and under the Unified QME we evaluate transitions rates from the excited states to the ground state at the frequency $\nu$,
$k=2 J(\nu) [n_B(\nu)+1]$.
Furthermore, the analytic solution presented next is general for other bath models, possibly with different forms for $k$.

Before proceeding to solve the dynamics, we first notice a symmetry that can be exploited to reduce the dimensionality of the problem:
We define $P(t)=\frac{1}{2}(\sigma_{22}(t)+\sigma_{33}(t))$ and obtain the new set of equations,
\bea
\dot{P} (t)&=&- k\sigma_{32}^R(t)-\phi P(t)+\frac{\phi-k}{2},
\label{eq:redP}
\eea
\bea
\dot{\sigma}_{32}^R(t)&=&-k\sigma_{32}^R(t)-\phi P(t)+\Delta\sigma_{32}^I(t)+\frac{\phi-k}{2},
\label{eq:red32r}
\eea
\bea
\dot{\sigma}_{32}^I(t)&=&-k\sigma_{32}^I(t)-\Delta\sigma_{32}^R(t).
\label{eq:red32I}
\eea
Here, for convenience we define the rate constant
\bea
\phi\equiv(1+2e^{-\beta \nu})k,
\eea
and used the normalization condition to eliminate $\sigma_{11}(t)$ from the EOM.

\subsubsection{Analytical solution for the dynamics with the LEPE method} 
\label{S1:Solution}

\begin{figure*}[hbtp!]
\includegraphics[width=2\columnwidth]{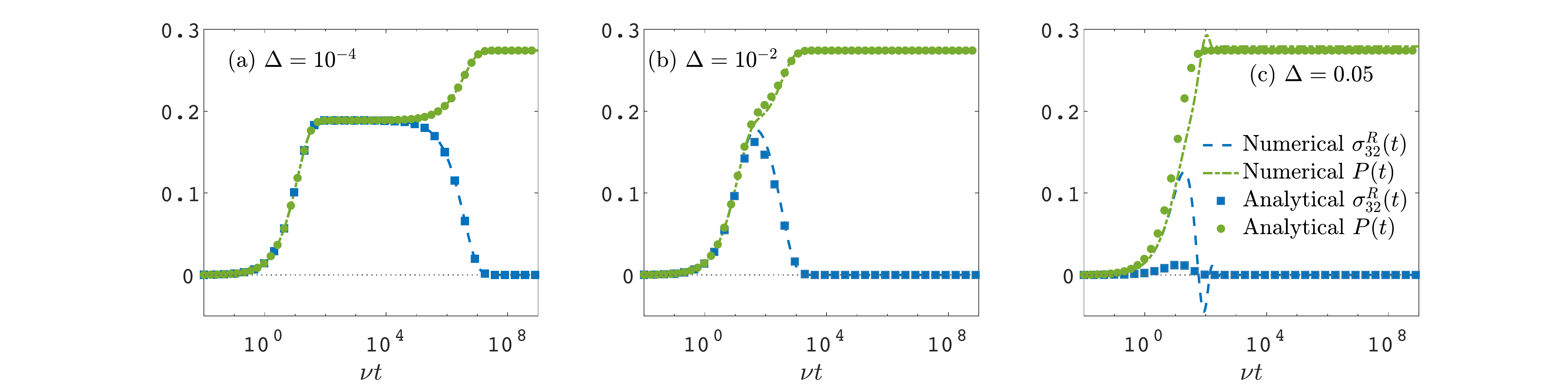} 
\caption{Dynamics of the excited state population $P(t)\equiv\frac{1}{2}[\sigma_{22}(t)+\sigma_{33}(t)]$ 
and real part of coherences, $\sigma_{32}^R(t)$ in the V model using the LEPE method with 
 the analytical solutions Eqs. (\ref{eq:Csol})-(\ref{eq:Psol}) (symbols). 
We compare these results to the numerical solution of the Redfield QME. 
The system is prepared in its ground state. 
Parameters are  $T=2$, $\gamma=0.005$; $\Delta$ is varied between the panels. Parameters are given relative to $\nu=1$.}
\label{fig:Fig1}
\end{figure*}

We solve here the dynamics of the V model, Eqs. (\ref{eq:redP})-(\ref{eq:red32I}), using the LEPE technique. 
We make the ansatz that the solution for each component of $\vec{x}(t)=(P(t), \sigma_{32}^R(t)$,  $\sigma_{32}^I(t))^{T}$ takes the form Eq. (\ref{eq:ansatz}).
There, $\lambda_n$ are the eigenvalues of the coefficient matrix, which is organized from Eqs. (\ref{eq:redP})-(\ref{eq:red32I}),
\bea
L = \underbrace{\begin{bmatrix}
-\phi & -k & 0 \\
-\phi & -k & 0\\
0 & 0 & -k\\
\end{bmatrix}}_{L^{(0)}}+ \underbrace{\Delta\begin{bmatrix}
0 & 0 & 0 \\
0 & 0 & 1\\
0 & -1 & 0\\
\end{bmatrix}}_{L^{(1)}}.
\label{L}
\eea
The other vector in Eq. (\ref{eq:MatrixQME}) is $\vec d = ( \frac{\phi-k}{2}, \frac{\phi-k}{2}, 0 )^{T}$. 

We also note that in the long-time limit, identified here by the $\infty$ symbol,
  $\sigma_{\infty,32}^R=\sigma_{\infty,32}^I=0$ and $P_{ \infty}=\frac{\phi-k}{2\phi}$; this is exactly the Gibbs state at the bath's temperature $T$.
  We thus put together the steady state vector
 $\vec x_\infty = (\frac{\phi-k}{2\phi}, 0 ,0)^{T} $.
While the system eventually equilibrates to the canonical state, its transient coherent dynamics can be very long-lived for nearly degenerate systems since (as we now show), $\lambda_1\propto {\Delta^2}$. 
We shall readily uncover this behavior with the LEPE method.

The characteristic polynomial of $L$ is given by
\bea
f(\lambda) = \Delta^2(\phi+\lambda)+\lambda(k+\lambda)(k+\phi+\lambda)=0,
\label{CP}
\eea
with the solutions $\lambda_n$.
However, this is a cubic equation. Even though it is possible to solve this equation exactly with e.g., Cardano's Formula, in general this is a difficult task.
In contrast, the eigenvalues of $L^{(0)}$, $\lambda_n^{(0)}$, are readily available due to its block diagonal form, resulting in
\bea
\lambda_1^{(0)}=0,\,\,\,\,\,\,\, 
\lambda_2^{(0)}=-\phi-k,\,\,\,\,\,\,
\lambda_3^{(0)}=-k.
\eea
In the limit $\Delta\to0$, we now make the ansatz  $\lambda_n\approx\lambda_n^{(0)}+\delta_n$, with $\delta_n$ as small corrections
to $\lambda_n^{(0)}$. This ansatz holds as long as $\abs{\delta_n}\ll\abs{k},\abs{\phi}$.
Starting with the first eigenvalue, $n=1$, we substitute $\lambda_1^{(0)}+\delta_1$ into the characteristic polynomial Eq. (\ref{CP}) and extract $\delta_n$,
\bea
\lambda_1=\delta_1=-\frac{\phi\Delta^2}{k(k+\phi)} + {\cal O}(\Delta^3).
\label{CP1}
\eea
This process is repeated to find $\delta_2$ and $\delta_3$, 
\bea
\delta_2&=&\Delta^2\frac{\phi-k}{\phi k}+{\cal O}(\Delta^3) ,\label{del2}\\
\delta_3&=&-\Delta^2\frac{\phi-k}{\phi(k+\phi)}+{\cal O}(\Delta^3).
\label{del3}
\eea
The latter two corrections are manifestly small compared to $\lambda_{2}^{(0)},\lambda_{3}^{(0)}$, and thus they are ignored. Overall, we obtained the eigenvalues for $L$ (assuming nearly-degenerate excited states) as
\bea
\lambda_1\approx&-\frac{\phi\Delta^2}{k(k+\phi)},\,\,\, \lambda_2\approx-\phi-k,\,\,\, \lambda_3\approx-k,
\label{eq:lambda}
\eea
where we only keep the leading terms in the expansion.
Notice that the matrix $L$ is singular precisely when $\Delta=0$, resulting in steady-states solutions that depend on initial conditions. Here, we always operate in the regime where $\Delta$ is small, but never identically zero. To provide physical insights, we express the eigenvalues in terms of the temperature
substituting $\phi$ by $(1+2e^{-\beta \nu})k$,
\bea
\lambda_1\approx&\frac{-\Delta^2(1+2e^{-\beta\nu})}{2k(1+e^{-\beta\nu})}, \lambda_2\approx-2k(1+e^{-\beta \nu}), \lambda_3\approx-k. 
\nonumber\\
\label{eq:lambdaT}
\eea
Recall that the relaxation rate constant $k$ can further depend on temperature [see Eq. (\ref{eq:kk})], yet this dependence is weak.

Reflecting on our results:
The eigenvalues of the Liouvillian dictate the timescale for equilibration, and we  resolved them in Eq. (\ref{eq:lambda}) with minimal efforts. The first eigenvalue is significantly smaller than the other two, thus it dictates the long-time decay to equilibrium, which stretches as the inverse of $\Delta^2$. 
Significantly, this slow decay dynamics is {\it robust against temperature} with $\lambda_1\approx -\frac{\Delta^2}{2k}$
at zero temperature, $\beta\nu\gg 1$, and $\lambda_1\approx -\frac{3\Delta^2}{4k}$
at high temperatures, $\beta\nu\to 0$.

We proceed and relate the initial conditions for the density matrix elements to the coefficients $c_{n,i}$ using the process described in Eqs. (\ref{eq:Omega})-(\ref{eq:BB}). Practically, one needs to prepare the matrices and vectors in Eq. (\ref{eq:BB}) for each element of the reduced density matrix. This process is detailed in Appendix A. 

As an example, we start the dynamics from the {\it ground state}, with {\it no initial coherence},
 $\sigma_{32}^R(0)=0$,  $\sigma_{32}^I(0)=0$, $P(t=0)=0$. Solving 
 (\ref{eq:BB}) in the limit $\Delta\to 0$ we retrieve  the coefficients $c_{n,i}$ from matrix inversion.
Putting together these results into Eq. (\ref{eq:ansatz}), we get
\bea
\sigma_{32}^R(t)=\frac{\phi-k}{2(\phi+k)}\Big[e^{-\frac{\phi\Delta^2}{k(k+\phi)}t}-e^{-(\phi+k)t}\Big]. 
\label{eq:Csol} 
\eea
This solution is valid for an arbitrary temperature.
As such, it is \textit{distinct} from that of Ref. \cite{dodin_quantum_2016}; the expressions agree once we take here the limit $T\to0$. 

It is significant to note that we arrived at Eq. (\ref{eq:Csol}) with minimal efforts by simple algebraic manipulations. 
Since we work in the limit where $\Delta\to 0$, this result immediately reveals the existence of two separate timescales in the problem:
a short timescale $\tau_2=1/(k+\phi)$ that dictates the time necessary to build coherences in the system from the initial condition, and a long timescale, over which the quasi-stationary coherences survive, $\tau_1=\frac{k(k+\phi)}{\phi\Delta^2}$, before equilibration.
A quasi-equilibrium state exists at intermediate times, with quasi-stationary coherences 
\bea
\sigma_{23}^R(\tau_2\ll t\ll \tau_1 )= \frac{\phi-k}{2(\phi+k)} = \frac{e^{-\beta\nu}}{2(1+e^{-\beta \nu})}.\nonumber\\
\label{eq:sigL}
\eea
Notably, contrary to common intuition, coherences in this case are {\it maximized} in the {\it high} temperature limit ($\beta\to 0$), reaching the value of $1/4$.
To explain this finding, recall that Eq. (\ref{eq:sigL}) was derived assuming that the system  initially occupies only the ground state, with unoccupied excited states and no coherences, see text above Eq. (\ref{eq:Csol}). Coherences between  excited levels are then generated due to bath-induced excitations, and this process is enhanced when the temperature is increased. 

Note that a third timescale $\tau_3=\lambda_3^{-1}$ does not show up in the (approximate) dynamical solution, which only reflects the slowest timescale $\tau_1=\lambda_1^{-1}$ and the next-faster one, $\tau_2=\lambda_2^{-1}$.
In other models, the dynamics may reflect other faster processes as well.

We point out that it is appropriate to refer to this long-lived solution during $\tau_2\ll t\ll \tau_1$ as a quasi-equilibrium state, or a prethermalized state. This is because  the coherences [and similarly the population, Eq. (\ref{eq:PsolL})], depend only on the temperature and the system's energetic, and not on the relaxation rates $k$, similarly to a true equilibrium solution, which does not reveal the timescale it took to reach it.
However, the quasi-equilibrium solution (\ref{eq:sigL}) implicitly depends on the initial condition. That is, different preparations lead to different values for this quasi-stationary coherences. 

Interestingly, the quasi-degenerate V model thus realizes ideas discussed in the prethermalization literature, see e.g., \cite{Mori_2018, Ueda_2020, Rigol_2019, Gring2012}, but in the setting of a weakly-coupled open quantum system where quantum coherences build up and sustain. 
While in our work the perturbation  lifts levels' degeneracy, in prethermalization models it is typically a driving parameter that is responsible for the intermediate dynamics.

The procedure (\ref{eq:BB}) is repeated to solve for $P(t)$, see Appendix A, and we find that the excited state population is
\bea
P(t)=\frac{\phi-k}{2\phi}-\frac{\phi-k}{2(\phi+k)}\Big[\frac{k}{\phi}e^{-\frac{\phi\Delta^2}{k(k+\phi)}t}+e^{-(\phi+k)t}\Big],
\nonumber\\
\label{eq:Psol}
\eea
starting the evolution from zero population in the excited states.
The quasi-equilibrium population is given by 
\bea
P(\tau_2\ll t\ll \tau_1 )= \frac{\phi-k}{2(\phi+k)} = \frac{e^{-\beta\nu}}{2(1+e^{-\beta \nu})},\nonumber\\
\label{eq:PsolL}
\eea
which is smaller than the equilibrium population by a factor $\phi/(\phi+k)$. 
Interestingly, for the ground-state initial condition, the ``prethermalized" population equals coherences, see Eq. (\ref{eq:sigL}).
As for the imaginary part of the reduced density matrix, 
the three coefficients $c_{n,3}$ are order of $\Delta$
and therefore we neglect them, and from henceforth focus on the dynamics of the population and the real part of coherences only.


\subsubsection{Simulations}

Until this point, the properties of the thermal bath were not used in calculations: The solution to the dynamics is the same whether the bath is fermionic or bosonic, harmonic or anharmonic [see Eq. (\ref{eq:HB})], and what particularly the operator $\hat B$ in Eq. (\ref{eq:Vm}) is. 

To perform simulations, we need to specify the rate in Eq. (\ref{eq:kk}). We
assume that the bath is harmonic and that $\hat B$ is a displacement operator.
The spectral density of the bath is assumed ohmic $J(\omega)=\gamma\omega e^{-\omega/\omega_c}$ with a very high frequency cutoff $\omega_c$. 
The dimensionless prefactor $\gamma$ controls the strength of the system-bath coupling, and it should be made small, $\gamma\ll1$, for ensuring the dynamics is consistent with the underlying weak-coupling approximation. 
Note that in the weak coupling limit, the  functional form of the spectral density function carries little impact, since rate constants are calculated at the specific transition frequency. 
Furthermore, to be able to exercise the LEPE approach we assume that $J(\omega)$
changes slowly enough with frequency around $\nu$ such that $J(\nu)\simeq J(\nu-\Delta)$.
This assumption holds for generic spectral functions, as long as they do not support sharp peaks (with peak width order of or smaller than $\Delta$) around the gap frequency $\nu$.


We exemplify the dynamics of the V model in Fig. \ref{fig:Fig1} by presenting
the real part of the coherences and the levels' population. We find that  
Eqs. (\ref{eq:Csol}) and (\ref{eq:Psol}) provide excellent analytical approximations to the dynamics, 
obtained numerically from the Redfield QME. The agreement holds from short time to equilibrium---so long as we work in the appropriate limit of $\Delta\ll k$
 and $\Delta \ll \nu$, underlying the Unified QME, and the LEPE method.

First, in Fig. \ref{fig:Fig1} (a) we display the coherences and populations for $\Delta/\nu = 10^{-4}$, which is well within the range of validity of the LEPE method. Indeed, we obtain a perfect agreement between the numerical Redfield solution and analytical expressions provided by LEPE. Both coherences and the population show long transient dynamics corresponding to the smallest eigenvalue $\lambda_1$.  In Fig. \ref{fig:Fig1} (b), we test the case $\Delta/\nu = 10^{-2}$. Here, the numerical Redfield
dynamics is mostly in agreement with our analytic expressions, with minor deviations showing up in the transient regime.
 As expected, coherences exist for a shorter duration, in accordance with Eq. (\ref{eq:lambda}). 
 Finally, in Fig. \ref{fig:Fig1} (c), $\Delta$ is made even bigger, resulting in both quantitative and qualitative deviations of the analytical expressions from the  numerical solutions. This is due to both
 the Unified QME becoming less accurate, as well as the LEPE method breaking down.


\section{Mpemba Effect in quantum dissipative Dynamics}
\label{S2:Mpemba}

\subsection{Principles of anomalous dynamics}

The ansatz for the dynamics of an open Markovian quantum system with quasi-degenerate levels, 
Eq. (\ref{eq:ansatz}), along with the  LEPE approach to find the corresponding eigenvalues and expansion coefficients, provide a natural platform to discuss relaxation towards equilibrium. 
In the weak system-bath coupling regime an open quantum system thermalizes to the Gibbs state in the long time limit. 
In Eq. (\ref{eq:ansatz}), the  $x_{\infty,i}$ terms are the values predicted by the Gibbs state, while the other terms in the expansion may be referred to as the decay channels, responsible for deviations from the equilibrium state in the transient regime. While each mode decays at its own timescale, $\tau_n = \frac{1}{|\lambda_n|}$, the timescale to thermalize is determined by the mode with the smallest eigenvalue in magnitude. We order our eigenvalues in increasing order, $|\lambda_n| < |\lambda_{n+1}|$;  the slowest mode is denoted by $\tau_1=\frac{1}{|\lambda_1|}$.

The expansion in Eq. (\ref{eq:ansatz}) for each element $x_i$ is given in terms of the set of coefficients  $c_{n,i}$, related to the initial conditions as prescribed by Eq. (\ref{eq:BB}).
In the standard approach to dynamics, which we exercised in Sec. \ref{S1:Model}, we assumed an initial condition for the  reduced density matrix, then resolved the $c_{n,i}$ prefactors. However, we may also approach the problem differently, enforce $c_{1,i} = 0$, $\forall i$ and translate this constraint to initial conditions on the elements of the reduced density matrix.
By excluding all the $c_{1,i}$ from the dynamics, we accelerate the equilibration process since its timescale is now dictated by 
$\tau_2 = \frac{1}{|\lambda_2|} <\tau_1$. 
Given this control over the equilibration process, it becomes possible to observe a quantum analog of the Mpemba effect:
We can initialize the system in separate initial states, but the one 
starting further away from equilibrium can reach the equilibrium Gibbs state {\it before} the other that begins {\it closer}, due to the former missing the small (thus long-lived) $\lambda_1$ decaying mode.  


\subsection{Mathematical analysis}

We return to the V model, Sec. \ref{S1:Solution} and exemplify the Markovian Mpemba effect as the system relaxes towards equilibrium.
Using Eq. (\ref{eq:BB2}), rather than dictating initial conditions for the elements of the reduced density matrix, 
we suppress the slow dynamics of the real part of the coherences by 
setting $c_{1,2} = 0$. Recall that the first index ($n$)  identifies the eigenvalue; the second index $i$ selects the element of the
reduced density matrix, which are organized here as $\vec x(t) =(P(t), \sigma_{32}^R(t), \sigma_{32}^I(t))^T$.
This condition is equivalent to preparing an initial state, and thus a subsequent evolution, which is orthogonal to the mode $n=1$.
We refer to such solutions as ``Mpemba states".

Under this condition, we solve the linear problem (\ref{eq:BB2}) and get
\bea
\sigma_{32}^R(0)&=&c_{2,2}+c_{3,2}, 
\nonumber\\
P(0)&=&\frac{\phi-k}{2\phi}+c_{2,2}+{\frac{c_{3,2} k}{k-\phi}-\underbrace{\frac{\Delta^2(c_{2,2}+c_{3,2})}{(k-\phi)\phi}}_{\to0}},
\nonumber\\
\sigma_{32}^I(0)&=&{\frac{c_{3,2}k\phi}{\Delta(k-\phi)}-\underbrace{\Delta\frac{c_{2,2}+c_{3,2}}{k-\phi}}_{\to 0}}.
\eea
To be consistent with the approximation $\Delta\to0$, 
the regularity of the term $\frac{c_{3,2}k\phi}{\Delta(k-\phi)}$ enforces $c_{3,2}$ to be on the order of $\mathcal{O}(\Delta)$ or less. For simplicity, henceforth we set $c_{3,2}$ to zero. 

We observe an interesting relationship between the initial coherences and populations,
\bea
\sigma_{32}^R(0)=
P(0)
-\underbrace{\frac{\phi-k}{2\phi}}_{P_{\infty}}
-{{
\underbrace{
\frac{\Delta\sigma_{32}^I(0)}{k} }_{\to 0}} }.
\eea
According to this expression,  the initial coherences in Mpemba states exactly match the difference between the initial and final populations.
Pictorially, coherences thus serve as a ``storage" space for the equilibrium state, and from there it can quickly take over the initial conditions.

Repeating the calculation of Eq. (\ref{eq:BB2}) for the populations of the V model, 
we find that $c_{3,1}\approx 0$ and that $c_{2,1}=c_{2,2}$, referred from now on as $c_2$.
Altogether, we construct the Mpemba ($M$) initial state as
\bea
&&{\sigma_M(0)}=
\nonumber
\begin{bmatrix}
1-2(\frac{\phi-k}{2\phi}+c_{2}) & 0 & 0 \\
0 & \frac{\phi-k}{2\phi}+c_{2} & c_{2}\\
0 & c_{2} & \frac{\phi-k}{2\phi}+c_{2}\\
\end{bmatrix}.
\\
\label{eq:initialstates}
\eea
This reduced density matrix is written in the original $|1\rangle$, $|2\rangle$, $|3\rangle$ basis of the V model.
The single free  coefficient $c_{2}$ must be chosen such that the reduced density matrix is physical, obeying the normalization, positivity, and purity conditions.
These additional constraints are distilled into the bounds 
\bea
\frac{3k-\phi-\sqrt{(5\phi-3k)(\phi+k)}}{8\phi}\leq c_2\leq\frac{k}{2\phi}, 
\label{eq:req0}
\eea
corresponding to taking  $-0.2431\lesssim c_2\lesssim 0.226$ with the parameters we use in this work ($T=2,\gamma=0.005,\nu=1$).
Evolving from this initial condition, the dynamics satisfies
\bea
\sigma_{M,32}^{R}(t)&=&
c_2e^{-(\phi+k)t}, \,\,\,\,\, \sigma_{M,32}^{I}(t)= 0,
\label{eq:Mpemba_coherence}
\nonumber\\
P_M(t)&=&c_2e^{-(\phi+k)t}+\frac{\phi-k}{2\phi}.
\eea
The $M$ subscript highlights that these expressions evolve from a Mpemba initial state, missing the slowest eigenmode.

\begin{figure}[hbt!]
\centering
\includegraphics[width=1\columnwidth]{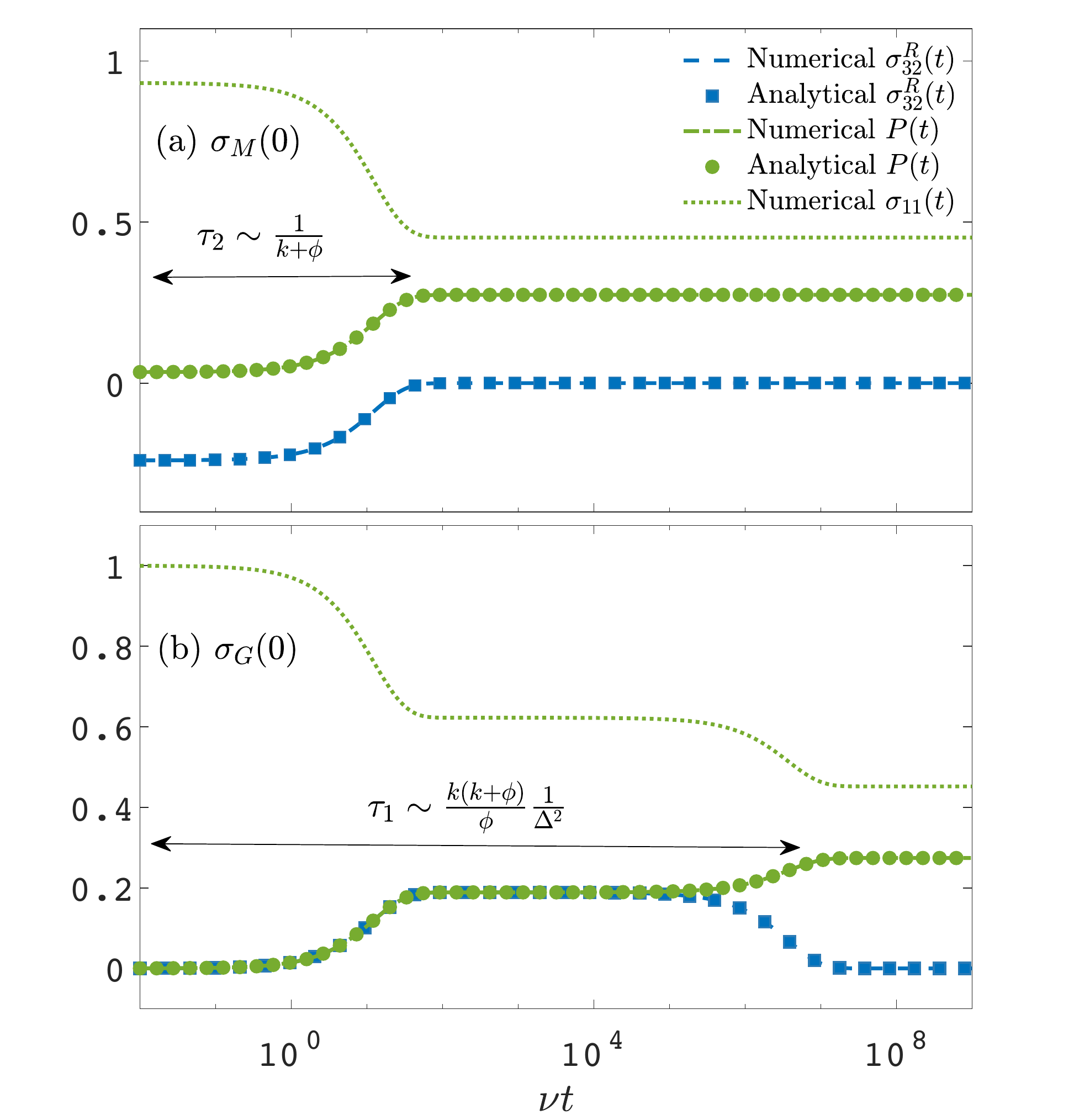} 
\caption{Accelerating equilibration dynamics of the V model. 
(a) Quantum dynamics from a Mpemba initial state $\sigma_M(0)$, Eq. (\ref{eq:initialstates}), with $c_{2}=-0.24$.
(b) Quantum dynamics from the ground state $\sigma_G(0) =|1 \rangle\langle 1|$.
 Analytical results were obtained directly from Eqs.  (\ref{eq:Csol})-(\ref{eq:Psol}) and (\ref{eq:Mpemba_coherence}), which rely on the LEPE method.
Numerical simulations were performed using the Born-Markov Redfield quantum master equation, with no approximations on the decaying eigenvalues.
Parameters are  $T=2$, $\gamma=0.005$, and $\Delta = 10^{-4}$, $\nu=1$.}
\label{fig:Mpemba}
\end{figure}

\subsection{Simulations}

Fig. \ref{fig:Mpemba} contrasts the dynamics of a V system initially prepared in a Mpemba state, $\sigma_M(0)$,
to the dynamics evolving from the ground state,  $\sigma_G(0) = |1\rangle\langle1|$.
We observe a dramatic difference in relaxation times under these two initial states, extending 5 orders of magnitude: The equilibration timescale is roughly $\nu \tau_2 \approx 20$ for the Mpemba state in Fig. \ref{fig:Mpemba}(a) and $\nu \tau_1\approx 3\times 10^6$ for the ground state preparation in Fig. \ref{fig:Mpemba}(b). 
A Mpemba state allows an acceleration of the relaxation dynamics, here by many order of magnitudes roughly dictated by the ratio $(k/\Delta)^2$.
The Mpemba effect is extreme in the V model since 
the slow dynamics is especially long-lived due to the presence of quasi-degenerate levels.
To quantify the equilibration dynamics we employ the following trace-distance as a distance-to-equilibrium measure, 
\bea
{\mathcal D}(\sigma_a,\sigma_b)=\frac{1}{2}\text{Tr}\Big[\sqrt{(\sigma_a-\sigma_b)^2}\Big].
\label{eq:trd}
\eea
Here, $\sigma_a$ and $\sigma_b$ are reduced density matrices. Crucially, this choice constitutes a physical  distance measure: It is a monotonically decreasing function under a CPTP map as the system relaxes towards equilibrium \cite{Raz}. 
Furthermore, the trace distance is also a metric on the space of density matrices, making it an appropriate choice to quantify the degree of closeness between states towards thermal equilibrium. 

\begin{figure}[hbt!]
\centering
\includegraphics[width=1\columnwidth]{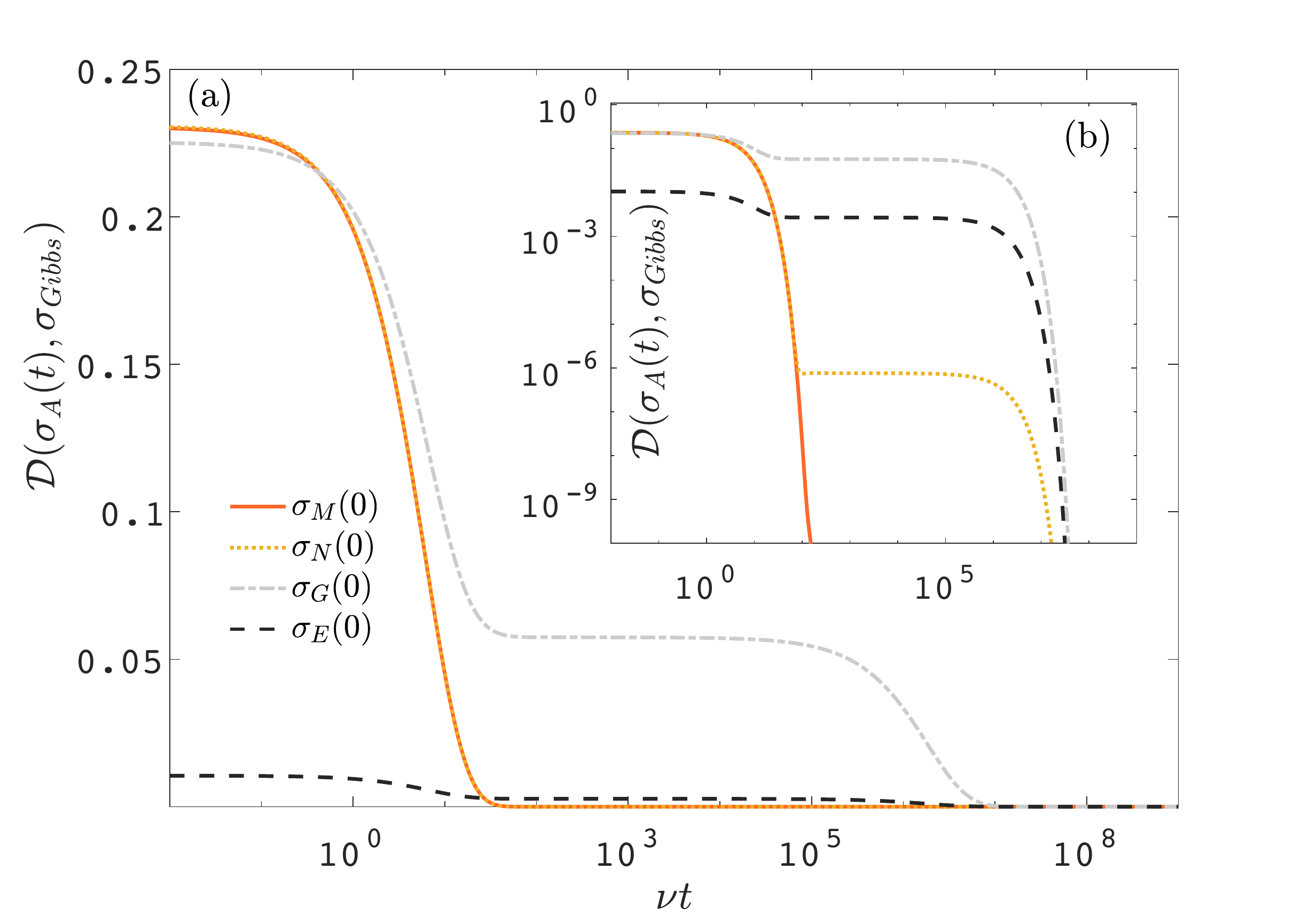} 
\caption{Trace distance computed with Eq. (\ref{eq:trd}), between the equilibrium Gibbs state of the V model, defined by $\sigma_{Gibbs}=\frac{e^{-\beta\hat{H}_S}}{\Tr[e^{-\beta\hat{H}_S}]}$ and $\sigma_A(t)$, evolving from:
a Mpemba initial state $\sigma_M(0)$ with $c_2=-0.24$ (full), a perturbed Mpemba state, $\sigma_N(0) = \sigma_M(0) + (0.001c_{2}|2\rangle \langle 3| + h.c.)$ (dotted, yellow), the ground state of the V system $\sigma_G(0) = |1\rangle\langle 1|$ (dashed-dotted gray), and the maximally-mixed state $\sigma_E(0) = \frac{1}{3}(|1\rangle\langle 1|+ |2\rangle\langle 2| +|3\rangle\langle 3|)$ (dashed, black).
Panel (b) presents the same results, but on a log-log scale.
Parameters are the same as in Fig. \ref{fig:Mpemba}. 
}
\label{fig:TrD}
\end{figure}

In Fig. \ref{fig:TrD} we present the time evolution of the trace distance from certain initial conditions (we test four)  to the equilibrium state of the V model. The four initial states are:
(i)  A Mpemba initial condition, $\sigma_M(0)$, which precludes the slowest eigenmode,
and  (ii) $\sigma_N(0) = \sigma_M(0) + (0.001c_2|2\rangle\langle3| + h.c.)$, a state close to the Mpemba initial state, possessing only a small deviation in the coherences.
We further test two diagonal initial conditions:
(iii)  A ground state preparation, $\sigma_G(0)$, and
(iv) the maximally mixed state, $\sigma_E(0) = \frac{1}{3}(|1\rangle\langle1| + |2\rangle\langle2| +|3\rangle\langle3|)$.

Comparing the relaxation dynamics to the Gibbs state from the Mpemba state $\sigma_M(0)$, to that from the ground state $\sigma_G(0)$ or the maximally-mixed state $\sigma_E(0)$, we find that while the Mpemba state starts further away from thermal equilibrium than the other two, it reaches it significantly faster.
We also compare the dynamics starting from $\sigma_M(0)$ to the dynamics from $\sigma_N(0)$. At short-intermediate times, $\sigma_N(0)$ outruns $\sigma_G(0)$ and $\sigma_E(0)$. However, it does not decay to equilibrium as quickly as the Mpemba state, but lingers due to the slowest decaying mode being (slightly) populated.
Nevertheless, we conclude that one does not need preparing the Mpemba state very precisely to attain the hyper-acceleration of the relaxation dynamics to a state very close to equilibrium.
As long as the population of the slow mode is significantly lower than that of the other modes, a Mpemba effect can be practically realized.

\section{Discussion}
\label{S:Discussion}

Analytical solutions for the dynamics of the V model were presented in Refs. \cite{Tscherbul_2014,Tscherbul_2015,Dodin_2016}, and recently for the $\Lambda$ model in Ref. \cite{Tscherbul_2022}. However, a few important remarks are in order: (i) Our derivation relies on a simpler, newly-developed Liouvillian eigenvalue estimation technique that can be readily applied to other models possessing nearly-degenerate states such as the
 $\Lambda$ model, as we discuss in Appendix B. 
(ii) The results presented in the above references are valid at low temperatures. Our expressions on the other hand hold for a broader temperature range, applicable to e.g., biological processes. While in principle our expressions are thermodynamically consistent even at zero temperature, the Redfield equation is no longer accurate at low temperature with the breakdown of Markovianity. 
(iii) Our approach is also amenable to \textit{nonequilibrium} conditions (Appendix B).

The generic mechanism underlying the Mpemba effect is the careful preparation of an initial state which is orthogonal to the slowest decaying mode.
Thus, such states achieve a factor of $\frac{|\lambda_2|}{|\lambda_1|}$ acceleration to the decay rate, with $|\lambda_2|>|\lambda_1|$.  This ratio hovers around unity for {\it classical} systems where coherences do not play a role (see Appendix C).
However, the acceleration of the dynamics arising in a model with {\it quantum coherences}, as presented in this work, is distinct:
Since only the slowest eigenvalue $\lambda_1$ depends on $\Delta$, while all other modes are {\it independent} of it, the acceleration factor depends on this parameter, which can be very small. Thus, once the slow mode is turned off, the equilibration dynamics is accelerated by many orders of magnitude. 
Concretely, while ``conventional" states in the V model decay to equilibrium after a characteristic timescale $\tau_1=\frac{k(k+\phi)}{\phi \Delta^2}$,  Mpemba states decay faster, after $\tau_2=\frac{1}{(k+\phi)}$, {\it irrespective} of $\Delta$ -- the acceleration factor is thus $\frac{k(k+\phi)^2}{\phi \Delta^2}$,  or roughly $(k/\Delta)^2$.

To contrast the quantum Markovian Mpemba effect, which is enacted by the control of coherences,
to the secular-incoherent limit, we study the latter case in Appendix C. We show there that in the classical regime we lack the strong $\Delta$ tunability underpinning the hyper-acceleration towards equilibrium. 


\section{Conclusion}
\label{S3:Summary}

We presented a Liouvillian eigenvalue perturbative estimation technique, an analytic approach to feasibly solve the dissipative dynamics of open Markovian quantum systems with quasidegenerate levels. This LEPE approach relies on decomposing the dissipator into the solvable (degenerate) part and the perturbative part lifting degeneracies.
Using the V model as a case study, we demonstrated that the LEPE method was accurate when the excited states were close to degeneracy. Specifically, the LEPE method allows a straightforward identification of relaxation timescales in the equilibration process.
We further explained how to construct Mpemba-like states in the V model. These states  bypass the slow bath-induced coherent dynamics. We found that in the V model, Mpemba states should be prepared with quantum coherences between excited states exactly matching the difference between final and initial populations of the excited states.
With such a preparation, the system reached equilibrium in an hyper-accelerated time. Namely, the system bypassed long-lived coherent dynamics with lifetime ${\cal O}(\frac{k}{\Delta^2})$, and instead relaxed after time that was independent of $\Delta$. There are many Mpemba states; in the V model they lie in the region defined by Eq. 
(\ref{eq:req0}). Furthermore, even sub-optimal Mpemba states with residual weights in the slow mode  display accelerated decay to equilibrium, compared to standard preparation (Fig. \ref{fig:TrD}).



The LEPE method can be applied to more general Markovian dynamics than examined here, such as when the system
is coupled to multiple heat baths (Appendix B)
or when many quasidegenerate levels exist in the spectrum.
While demonstrated here on the Unified quantum master equation, which is derived from the Redfield equation but is a CPTP map and thermodynamically consistent,   the LEPE method can be utilized in other cases, such as when the EOMs include higher-order system-bath coupling effects as in the polaron-transformed QME.

Thermal baths can lead to long-lived transient coherences in a quantum system, with delayed equilibration, which could impact e.g., quantum thermometry \cite{thermometry}.
One can however bypass this slow dynamics by initializing the system in a way that cuts off the slow modes. Future work will be focused on using the LEPE method to uncover the equilibration timescale of systems experiencing strong couplings to their environment.

\begin{acknowledgments}
DS acknowledges the NSERC discovery grant and the Canada Research Chair Program. 
NAS acknowledges support from the Ontario Graduate Scholarship.
The work of FI was funded by the University of Toronto Excellence Award.
The authors acknowledge John Bechhoefer for introducing us to the classical Mpemba effect, and Lianao Wu for 
discussions on the quantum Mpemba effect.
\end{acknowledgments}

\begin{widetext}
\renewcommand{\theequation}{A\arabic{equation}}
\setcounter{equation}{0}  
\setcounter{section}{0} 
\section*{Appendix A: The LEPE method on the V model: Resolving the coefficients of the eigenmodes}
\label{app:A}

We demonstrate on the V model how to relate initial conditions on the reduced density matrix to the mode coefficient matrix.
We begin with $\sigma_{32}^R(t)$, which is the second element in $\vec x(t)$. 
Explicitly, it evolves as
\bea
x_2(t)=x_{\infty,2} + c_{1,2} e^{\lambda_1t}
+c_{2,2} e^{\lambda_2t}+c_{3,2} e^{\lambda_3t}.
\eea
Our goal is to express $c_{n,2}$ in terms of the initial conditions. The three decay rates $\lambda_n$ were already obtained with the LEPE method, and they are given by Eq. (\ref{eq:lambda}).

We customize Eq. (\ref{eq:BB}) as
\bea
B^{(2)} \vec x(0) + \vec v^{(2)}=\Lambda \vec c^{(2)} +\vec x_{\infty}^{(2)}.
\label{eq:BB2}
\eea
 The right hand side is defined based on the relation
 $\vec{s}_{32}^R(0)=\Lambda \vec c^{(2)}(0) + \vec x_{\infty}^{(2)}$, with
%
%
\bea
\vec{s}^R_{32}(0)&=&
\begin{bmatrix}
\sigma_{32}^R(0)\\
\dot \sigma_{32}^R(0)\\
\ddot\sigma_{32}^R(0)\\
\end{bmatrix}, \,\,\,\,\,\,\,
\Lambda=
\begin{bmatrix}
1 & 1 & 1 \\
\lambda_1 & \lambda_2 & \lambda_3\\
\lambda_1^2 & \lambda_2^2 & \lambda_3^2\\
\end{bmatrix},\,\,\,
\vec{c}^{(2)}=
\begin{bmatrix}
c_{1,2}\\
c_{2,2}\\
c_{3,2}\\
\end{bmatrix},\,\,\,\,\,\,
\vec{x}_{\infty}^{(2)}=
\begin{bmatrix}
0\\
0\\
0\\
\end{bmatrix}.
\eea
The left hand side of (\ref{eq:BB2}) is constructed from the matrices and vectors
\bea
B^{(2)} &=&
\begin{bmatrix}
0 & 1 & 0 \\
-\phi & -k & {\Delta} \\
 \phi(k+\phi) & k(k+\phi)-\Delta^2 &-2\Delta k\\
\end{bmatrix}, 
\nonumber\\
\vec{x}(0)&=&
\begin{bmatrix}
P(0)\\
\sigma_{32}^R(0)\\
\sigma_{32}^I(0)\\
\end{bmatrix}, \
\vec{v}^{(2)}=
\begin{bmatrix}
0\\
\frac{\phi-k}{2}\\
\frac{k^2-\phi^2}{2}\\
\end{bmatrix}.
\eea
%

We decide to start the dynamics from the the {\it ground state}, with {\it no initial coherence},
 $\sigma_{32}^R(0)=0$,  $\sigma_{32}^I(0)=0$, $P(0)=0$. Solving  Eq.
 (\ref{eq:BB2}) in the limit of $\Delta\to 0$ we get the coefficients $c_{n,2}$ from matrix inversion,
\bea
c_{1,2}=\frac{\phi-k}{2(\phi+k)},\,\,\,\, \,\,c_{2,2}=-\frac{\phi-k}{2(\phi+k)}, \,\,\,\,\,\,
c_{3,2}=-\frac{\Delta^2(\phi-k)}{2k^2(\phi+k)}\approx0.
\eea
The procedure is repeated to solve for $P(t)$. 
In this case,
\bea
B^{(1)} &=&
\begin{bmatrix}
1 & 0 & 0 \\
-\phi & -k & 0 \\
 \phi(k+\phi) & k(k+\phi) &-\Delta k\\
\end{bmatrix}, \,\,\,\,\,\,\,\,
\vec{v}^{(1)}=
\begin{bmatrix}
0\\
\frac{\phi-k}{2}\\
\frac{k^2-\phi^2}{2}\\
\end{bmatrix},\,\,\,\,\,\,\,\,\,
\vec{x}_{\infty}^{(1)}=
\begin{bmatrix}
\frac{\phi-k}{2\phi}\\
0\\
0\\
\end{bmatrix}.
\eea
Based on the specific initial condition of zero initial excited population and coherences, we find the coefficients
\bea
c_{1,1}=-\frac{k(\phi-k)}{2\phi(\phi+k)},\,\,\,\, \,\,c_{2,1}=-\frac{\phi-k}{2(\phi+k)}, \,\,\,\,\,\,
c_{3,1}=\frac{\Delta^2(\phi-k)}{2k\phi(\phi+k)}\approx0.
\eea
As for $\sigma_{32}^I$, the analysis results in $c_{n,3}={\cal O}(\Delta)$ for $n=1,2,3$, thus the three coefficients are neglected.

\renewcommand{\theequation}{B\arabic{equation}}
\renewcommand{\thesubsection}{B\arabic{subsection}}
\setcounter{equation}{0}  
\setcounter{section}{0} 
\setcounter{subsection}{0} 
\section*{Appendix B: Analogous models and extensions}
\label{app:2}

\subsection{The $\Lambda$ model}

The $\Lambda$ model includes one excited level ($|3\rangle$) and two levels of lower energy, $|1\rangle$ and $|2\rangle$.
The Hamiltonian of the model is
\bea
\hat{H}_S^\Lambda=\Delta|2\rangle\langle2|+\nu|3\rangle\langle3|,
\eea
and we assume that levels $|1\rangle$ and $|2\rangle$ are nearly degenerate with the energy of level $|1\rangle$ set to zero and $\Delta\ll\nu$.
The  heat bath allows excitations from the two lowest levels to the excited state,
\bea
\hat{S}^\Lambda=|1\rangle\langle3|+|2\rangle\langle3|+h.c.
\eea
Similarly to the V model, we derive the Unified QME.
For the present model it is given by
\bea
\dot{\sigma}_{12}(t)&=& -i\Delta\sigma_{12}(t)
-
k e^{-\beta\nu}\sigma_{12}(t)
- \frac{1}{2}
ke^{-\beta\nu} \left[ \sigma_{11}(t)+  \sigma_{22}(t) \right] + 
k \sigma_{33}(t),
\label{Aeq:s23} 
\nonumber\\
\dot{\sigma}_{11}(t)&=&
- ke^{-\beta \nu} \sigma_{11}(t) 
+ k\sigma_{33}(t) - 
ke^{-\beta \nu} \sigma_{12}^R(t),
\label{Aeq:s22}
\nonumber\\
\dot{\sigma}_{22}(t)&=&
-
ke^{-\beta \nu} \sigma_{22}(t) + 
k \sigma_{33}(t) - 
ke^{-\beta \nu} \sigma_{12}^R(t).
\label{Aeq:s33}
\eea
By replacing here $k$ by $ke^{-\beta\nu}$, identifying  $P(t)\equiv\frac{1}{2}(\sigma_{11}(t)+\sigma_{22}(t))$ and $\phi\equiv(2+e^{-\beta\nu})k$,  we recover Eqs. (\ref{eq:redP})-(\ref{eq:red32I}), with $\sigma_{23}\leftrightarrow\sigma_{12}$. As such, 
all the results derived in the paper generalize to the $\Lambda$ model.

\subsection{Nonequilibrium models}
Consider the equations of motion of the V model, but coupled to $N$ reservoirs held at different temperatures. Each bath
enacts the same transitions from the ground state 
$|1\rangle$ to the excited levels,
\bea
\hat{S}_1=\hat{S}_2=...=\hat{S}_N=|1\rangle\langle2|&+&|1\rangle\langle3|+h.c.
\eea
Otherwise, the same Hamiltonian is used as in the main text.
The equations of motion for the reduced density matrix, resulting directly from the additivity of the Redfield master equation, are given by the generalization of Eqs. (\ref{eq:s23})-(\ref{eq:s33}),
\bea
\dot{\sigma}_{32}(t)&=& -i\Delta\sigma_{32}(t)
-
\sum_j^N k_j\sigma_{32}(t)+
\sum_j^N k_j e^{-\beta_j\nu} \sigma_{11}(t)
- \frac{1}{2}
\sum_j^N k_j \left[ \sigma_{22}(t)+  \sigma_{33}(t) \right],
\\
\dot{\sigma}_{22}(t)&=&
- \sum_j^N k_j \sigma_{22}(t) 
+ \sum_j^N k_je^{-\beta_j \nu}\sigma_{11}(t) 
- 
\sum_j^N k_j \sigma_{32}^R(t),
\\
\dot{\sigma}_{33}(t)&=&
-
\sum_j^N k_j \sigma_{33}(t) + 
\sum_j^N k_j e^{-\beta_j \nu} \sigma_{11}(t) 
- 
\sum_j^N k_j \sigma_{32}^R(t).
\eea
By defining $\phi\equiv\sum_j\phi_j$ and $k\equiv \sum_jk_j$, we recover 
Eqs. (\ref{eq:s23})-(\ref{eq:s33}), which then transform as in the main text to the form Eqs. (\ref{eq:redP})-(\ref{eq:red32I}). 
As such, our results in the paper generalize to the nonequilibrium multi-bath case.

\subsection{Asymmetric models}

The LEPE approach provides a natural platform to study the transient dynamics of the modified V model, recently investigated in Ref. \cite{NJP}. In this setup, the V system is coupled to two baths held at different temperatures, but in an asymmetric manner enacted by an additional $\alpha$ parameter, see Ref. \cite{NJP} for the explicit definition and full Hamiltonian. 
The $\alpha$ parameter controls the interference behavior of the model  affecting the dynamics and the steady state results.

In Ref. \cite{NJP}, our focus has been on the steady state heat transport trends in this $\alpha$ interference-controlled V model.
To study the dynamics of the model,
the procedure of Sec. \ref{S1:Model} may be undertaken. 
Doing so we conclude that the lifetime of coherent dynamics (and subsequently the extent of Mpemba acceleration) is long-lived only near $\alpha=1$. Details will be presented in a future publication. 


\renewcommand{\theequation}{C\arabic{equation}}
\setcounter{equation}{0}  
\setcounter{section}{0} 
\section*{Appendix C: Incoherent Mpemba decay dynamics}
\label{app:1}

The role of quantum coherences in the V model dynamics, as examined in the main text, is contrasted here to its classical behavior.
This is done by considering the quantum master equation of the V model, Eqs. (\ref{eq:s23}) - (\ref{eq:s33}), but
 in its fully secular limit where we ignore terms coupling coherences and populations. 
Furthermore,  we do not enforce here the excited levels to be nearly degenerate, thus we maintain the rates $k_{1\to 2}$ and $k_{1\to 3}$ distinct. Here, $k_{1\to 3}$ is the excitation rate constant from level 1 to 3; $k_{1\to2}$ is the analogous rate of transitioning from level 1 to 2. The incoherent master equation for this three-level model (\ref{eq:Hs})-(\ref{eq:S}) is written as
\bea
    \dot{p}_1(t) &=& k_{3\to1}p_3(t) + k_{2\to1}p_2(t) - (k_{1\to3} + k_{1\to2})p_1(t),
    \nonumber\\
    \dot{p}_2(t) &=& -k_{2\to1}p_2(t) + k_{1\to2}p_1(t),
    \nonumber\\
    \dot{p}_3(t) &=& -k_{3\to1}p_3(t) + k_{1\to3}p_1(t).
\eea
Here, $p_j(t)$ denotes the population of level $j$.
Next, we make use of the normalization condition of the populations $1=\sum_jp_j(t)$ and the detailed balance relation between the rate constants, $k_{1\to3} = k_{3\to1}e^{-\nu\beta}$ and $k_{1\to2} = k_{2\to1}e^{-(\nu-\Delta)\beta}$, to transform the equations to a $2\times2$ system,
\bea
    \nonumber
    \dot p_2(t) &=& -k_{2\to1}\left(1 + e^{-(\nu-\Delta)\beta}\right)p_2(t)
    -k_{2\to1}e^{-(\nu-\Delta)\beta}p_3(t) + k_{2\to1}e^{-(\nu-\Delta)\beta},
   \nonumber\\
    \dot p_3(t) &=& -k_{3\to1}\left(1 + e^{-\nu\beta}\right)p_3(t) 
    - k_{3\to1}e^{-\nu\beta}p_2(t) + k_{3\to1}e^{-\nu\beta}.
\eea
We propose a solution of the form
\bea
    p_j(t) = p_{\infty,j} + c_{1,j}e^{\lambda_1t} + c_{2,j}e^{\lambda_2t},
    \label{eq:classpop}
\eea
where $p_{\infty,j}$ is the long-time population of the $j$th level, obeying classical Boltzmann statistics,
\bea
    p_{\infty,2} &=& \frac{e^{-(\nu-\Delta)\beta}}{1 + e^{-(\nu-\Delta)\beta} + e^{-\nu\beta}},
    \nonumber\\
   p_{\infty,3} &=& \frac{e^{-\nu\beta}}{1 + e^{-(\nu-\Delta)\beta} + e^{-\nu\beta}}.
\eea
%
\begin{figure}[hbt!]
\centering
\includegraphics[width=0.65\columnwidth]{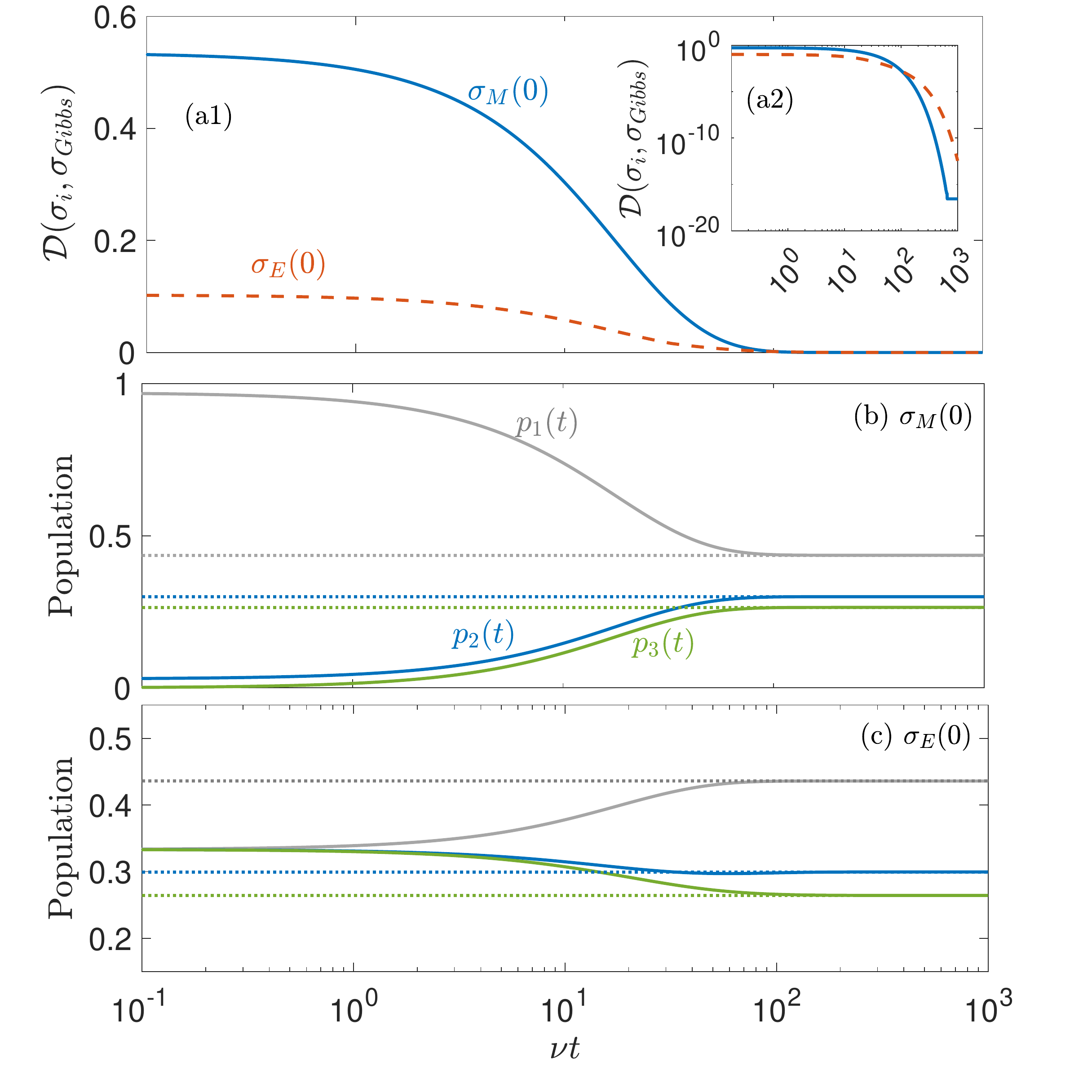}
\caption{Acceleration towards thermalization in the incoherent limit of the V model. 
(a1) The dynamics of the trace distance Eq. (\ref{eq:trd}) for a state initialized in a Mpemba state $\sigma_M(0)$ and the maximally mixed state $\sigma_E(0)$.
 (a2) Same quantities on a logarithmic scale, highlighting the deviations. 
 (b)-(c) The dynamics of the population given by Eq. (\ref{eq:classpop}) for the Mpemba state and the maximally mixed state, respectively. The long time values are presented by dotted lines.
 Parameters are $\nu = 1$, $\Delta = 0.25$, $T = 2$, $\gamma = 0.005$. 
 In panel (b), $c_{2,2} = -0.2703$ and $c_{2,3} = -0.2644$ while $c_{1,2}=c_{1,3}=0$.
 In panel (c), we begin with a maximally-mixed state.
 }
\label{fig:Secular}
\end{figure}
The Liouvillian is written as,
\bea
L =
\begin{bmatrix}
-k_{2\to1}[1 + e^{-(\nu-\Delta)\beta}] &-k_{2\to1} e^{-(\nu-\Delta)\beta}
 \\
-k_{3\to1}e^{-\nu\beta} & -k_{3\to1}(1 + e^{-\nu\beta})\\
\end{bmatrix}.
\eea
Here we are interested in the exact dynamics of the V model in the incoherent limit, thus we do not invoke any assumption on 
$\Delta$.
The eigenvalues $\lambda_1$ and $\lambda_2$ of $L$ are given exactly by 
\bea
\nonumber
    2\lambda_{1,2} &=& -\left[k_{2\to1}(1+e^{-(\nu-\Delta)\beta}) + k_{3\to1}(1+e^{-\nu\beta})\right] 
    \nonumber\\
    &\pm &
    \sqrt{\left[k_{2\to1}\left(1+e^{-(\nu-\Delta)\beta}\right) + k_{3\to1}(1+e^{-\nu\beta})\right]^2 - 4k_{2\to1}k_{3\to1}\left[1 + e^{-(\nu-\Delta)\beta} + e^{-\nu\beta}\right]},
    \nonumber\\
\eea
%
and we order the eigenvalues such that $|\lambda_2| > |\lambda_1|$. 
In the limit of high temperature, and assuming the decay rates are similar, we find that $\lambda_{1}=-k$
and $\lambda_2=-3k$. In the opposite low temperature limit, $\lambda_1=-k_{2\to 1}$ and $\lambda_2=-k_{3\to1}$.
While the two eigenvalues are distinct, their ratio $\lambda_2/\lambda_1$ is of order 1.
This is the most notable difference between the quantum case (\ref{eq:lambda}) and the incoherent limit: In the classical limit the eigenvalues are of a similar magnitude while in the quasi-degenerate quantum case the parameter $\Delta$ dictates the slow decay and thus it holds an exceptional control over transient timescales.

We turn now to the four coefficients $c_{1,2}, c_{2,2}, c_{1,3}, c_{2,3}$ and relate them to the initial level populations.
First, 
\bea
   p_j(0) &=& p_{\infty,j} + c_{1,j} + c_{2,j},
    \nonumber\\
    \dot{p}_j(0) &=& \lambda_1c_{1,j} + \lambda_2c_{2,j}. 
\eea
We focus as an example on the $j=2$ case.   
%
Similarly to the quantum case, we use the relation (\ref{eq:BB})
\bea
  B^{(2)}\vec{x}_0 + \vec{v}^{(2)} = \Lambda\vec{c}^{(2)} + \vec x_{\infty}^{(2)}.
  \label{eq:BB2A}
\eea
with the matrices and vectors,
\bea
B^{(2)}&=&
\begin{bmatrix}
1 & 0 \\
-k_{2\to1}(1+e^{-(\nu-\Delta)\beta}) & -
k_{2\to1}e^{-(\nu-\Delta)\beta} \\
\end{bmatrix}, \nonumber\\
\vec{x}_0&=&
\begin{bmatrix}
p_{2}(0)\\
p_{3}(0)\\
\end{bmatrix}, \,\,\,
\vec{v}^{(2)}=
\begin{bmatrix}
0\\
k_{2\to1}e^{-(\nu-\Delta)\beta}\\
\end{bmatrix}, \nonumber\\
\Lambda&=&
\begin{bmatrix}
1 & 1 \\
\lambda_1 & \lambda_2\\
\end{bmatrix}, 
\vec{c}^{(2)}=
\begin{bmatrix}
c_{1,2}\\
c_{2,2}\\
\end{bmatrix},\,\,\,
\vec{x}_{\infty}^{(2)}=
\begin{bmatrix}
p_{\infty,2}\\
0\\
\end{bmatrix}, \
\eea
With an algebraic manipulations we can now express $c_{n,2}$ in terms of the initial conditions, or vice versa. 
To analyze the Mpemba effect, we do the latter, obtaining
%
\bea
    p_2(0) &=& c_{1,2} + c_{2,2} + p_{\infty,2}
     \nonumber\\
    p_3(0) &=& - \frac{e^{(\nu-\Delta)\beta}}{k_{2\to1}}\Bigg\{\left[\lambda_1 + k_{2\to1}\left(1+e^{-(\nu-\Delta)\beta}\right) \right]c_{1,2}
    + \left[\lambda_2 + k_{2\to1}\left(1+e^{-(\nu-\Delta)\beta}\right) \right]c_{2,2} 
     \nonumber\\
    &-& k_{2\to1}e^{-(\nu-\Delta)\beta} + k_{2\to1}\left(1+e^{-(\nu-\Delta)\beta}\right)p_{\infty,2}\Bigg\}.
\eea
These initial conditions are not all physical, since any initial condition must satisfy $0\leq p_j(0)\leq1$, as well as $p_2(0) + p_3(0) \leq1$.
Going forward, we identify the ``Mpemba" state as those with $c_{1,2} = 0$. Using these set of constraints we find two pairs of conditions that must be satisfied simultaneously by $c_{2,2}$ in order for the states to be physical
\bea
    -p_{\infty,2} \leq &c_{2,2}& \leq 1 - p_{\infty,2}
    \\
   \frac{-k_{2\to1}\left[1 + e^{-(\nu-\Delta)\beta}\right]p_{\infty,2}}{\lambda_2 + k_{2\to1}\left[1+e^{-(\nu-\Delta)\beta}\right]}  \leq &c_{2,2}& 
   \leq \frac{k_{2\to1} \left[e^{-(\nu-\Delta)\beta} - \left(1 + e^{-(\nu-\Delta)\beta}\right)p_{\infty,2}\right]}{\lambda_2 + k_{2\to1}\left[1+e^{-(\nu-\Delta)\beta}\right]}
    \\
    &c_{2,2}& \leq -\frac{k_{2\to1}p_{\infty,2}}{\lambda_2 + k_{2\to1}}
\eea
%
The physical values of $c_{2,2}$ are given by the intersection of the above inequalities.
We repeat this process of building (\ref{eq:BB2A}) and solving it  for $j=3$. We again enforce the
slowest mode to vanish by setting $c_{1,3}$
and use the physical constraints on $p_2(0)$ and $p_3(0)$ to identify valid values for $c_{2,3}$.

Overall, the solution for the accelerated dynamics, missing the $\lambda_1$ eigenvalue, is given by
\bea
p_{2,M}(t) = p_{\infty,2} + c_{2,2}e^{\lambda_2t},
\\
p_{3,M}(t) = p_{\infty,3} + c_{2,3}e^{\lambda_2t}.
\eea
In Fig. \ref{fig:Secular}, we contrast this Mpemba-type dynamics, starting from $\sigma_M(0)$, to the case where the system starts at infinite temperature with equal population at the three levels, denoted by $\sigma_E(0)$.
To quantify the distance from equilibrium, we display in Fig. \ref{fig:Secular}(a) the trace distance Eq. (\ref{eq:trd})  of these two dynamics in relation to the thermal equilibrium state, $\sigma_{Gibbs}$, which is the long-time solution of the dynamics.
We find that despite starting further from equilibrium, the Mpemba state preparation fully thermalizes somewhat faster that a system initialized to  
$\sigma_E$, due to the Mpemba state missing the slowest decay mode. However, this effect is subtle in this model and of no practical ramifications.
In Fig. \ref{fig:Secular}(b)-(c), we follow the populations as a function of time for the system initialized in $\sigma_M(0)$ and $\sigma_E(0)$, respectively. On this scale it is impossible to distinguish between the two dynamics.
The quantum coherent case illustrated in Fig. \ref{fig:Fig1}, in contrast, displays a hyper-acceleration of the relaxation 
process for the Mpemba state due to the participation, and control of coherences in the process.

\vspace{4mm}
\end{widetext}

\bibliographystyle{quantum}
\bibliography{refsM}

\begin{thebibliography}{100}
\bibitem{comment3}
The analytical method is presented here for a model with only one manifold of
  quasi-degenerate levels, plus other discrete levels (as in the V level
  scheme). In this case, there is only a single perturbative parameter, the
  small frequency splitting in the spectrum. However, the method can be
  generalized to situations with several separate manifolds of quasi-degenarate
  levels, with their energy splittings serving as the perturbation parameters.
\bibitem{comment}
It is in principle possible to make this procedure a controlled approximation
  in $\Delta$, i.e., $\lambda_n=\lambda_n^{(0)}+\sum \delta_j\Delta^j$.
  However, this is not necessary in this work, as we rely on the Unified QME
  \cite{Anton} where $k_{3\to1}\approx k_{2\to1}$ since $\Delta\to 0$.
\end{thebibliography}
\end{document}